\documentclass[superscriptaddress,prl,floats,amsmath,amssymb,aps,twocolumn,showpacs]{revtex4-1}
\usepackage{graphicx}
\usepackage{amsmath}
\usepackage{amssymb}
\usepackage{dcolumn}
\usepackage{color}

\begin{document}

\hyphenation{ALPGEN}
\hyphenation{EVTGEN}
\hyphenation{PYTHIA}

\title{
The quantum nature of the superconducting hydrogen sulfide at finite temperatures}
\author{Ying Yuan}
\affiliation{School of Physics, Peking University, Beijing 100871, P. R. China}
\author{Yexin Feng}
\affiliation{School of Physics and Electronics, Hunan University, Changsha 410082, P. R. China}
\author{Lifeng Bian}
\affiliation{Key Laboratory of Nanodevices and Applications, Suzhou Institute of Nano-Tech and Nano-Bionics, Chinese Academy of Sciences, Suzhou 215123, Jiangsu, P. R. China}
\author{Dong-Bo Zhang}
\affiliation{Beijing Computational Science Research Center, Beijing 100193, P. R. China}
\author{Xin-Zheng~Li}
\email{xzli@pku.edu.cn}
\affiliation{School of Physics, Peking University, Beijing 100871, P. R. China}
\affiliation{Collaborative Innovation Center of Quantum Matter, Beijing 100871, P. R. China}

\begin{abstract}
H$_3$S is believed to the most possible high-temperature superconducting ($T_{\text{c}}$) phase
of hydrogen sulfide at $\sim$200 GPa.
It's isotope substitution of hydrogen (H) by deuterium (D), however, shows an anomalous $T_{\text{c}}$
decrease of $\sim$100 K at 140 to 160 GPa, much larger than the Bardeen-Cooper-Schrieffer theory
prediction.
Using \textit{ab initio} path-integral molecular dynamics (PIMD), we show that the nuclear quantum
effects (NQEs) influence the structures of H$_3$S and D$_3$S differently at finite temperatures
and the interval when H$_3$S possesses the symmetric high $T_{\text{c}}$ structure while D$_3$S does
not is in agreement with, though their absolute values are lower than experiments.
This is consistent with an earlier theoretical study using the stochastic self-consistent
harmonic approximation method in descriptions of the nuclei at 0 K.
The remaining discrepancy can be substantially improved when the electronic structures are
calculated using a hybrid function.
%
%
Our study presents a simple picture to interpret the isotope dependent of $T_{\text{c}}$ and emphasizes
the quantum nature in the high-pressure hydrogen sulfide system.
\end{abstract}

\date{\today}

\maketitle
\section{I. Introduction}
A superconductor is a material which exhibits zero electric resistance under a transition temperature ($T$),
\textit{i.e.} $T_{\text{c}}$.
Ever since its discovery in mercury in 1911~\cite{onnes1911}, understanding the mechanism of this behavior and
seeking for high-$T_{\text{c}}$ superconductor have ranked among the biggest challenges in physics.
Stimulated by the isotope-dependence of $T_{\text{c}}$ observed in a wide range of superconductors,
Bardeen, Cooper, and Schrieffer (BCS) proposed the microscopic picture of electron-phonon coupling in their
seminal work in the 1950s~\cite{bardeen1957a,bardeen1957b}, which explains the mechanism of most
superconductors discovered by then.
In 1986, a record-high 133 K $T_{\text{c}}$ was found in copper oxide at ambient pressure~\cite{bednorz1986},
followed by a 164 K $T_{\text{c}}$ in a similar system under pressure~\cite{gao1994},
and the discovery of a series of alike superconductors~\cite{putilin1993,schilling1993,dai1995}.
At present, the superconducting nature for most of these later found superconductors remains unclear.
Consequently, it is customary to call the superconductors whose superconducting behaviors can be explained
by BCS theory as the conventional ones.
Since BCS theory gives a clear guide to design high  $T_{\text{c}}$ conventional superconductors,
(\textit{i.e.} a high density of states close to the Fermi level, a favorable combination of high-frequency
phonons, and a strong electron-phonon coupling), metallic hydrogen and hydrogen-rich compounds were naturally
chosen as the target~\cite{ashcroft1968,ashcroft2004}.
Accordingly, a series of theoretical studies on hydrogen and hydrides had been carried
out (SiH$_4$\cite{ashcroft2004}, SnH$_4$\cite{ashcroft2004}, GeH$_3$\cite{Abe2013},
GeH$_4$\cite{Gao2008}, KH$_6$\cite{Zhou2012}, CaH$_6$\cite{Wang2012}, etc).
The crystal structure searching methods had played a crucial role~\cite{Pickard2006, Wang2012, Gao2010}.
For example, a high $T_{\text{c}}$ value of 98 to 107 K was predicted in SiH$_4$(H$_2$)$_2$ at
250 GPa (using a Coulomb parameter $\mu^*=0.1-0.13$, hereinafter,
see references for the details)~\cite{Li2010}.
In GeH$_4$(H$_2$)$_2$, a $T_{\text{c}}$ value of 76 to 90 K was predicted at 250 GPa~\cite{Zhong2012}.
In MH$_3$ (M$=$ Ga, Ge, Si) hydrides, $T_{\text{c}}$ values scattering from 76 K to 153 K were
reported~\cite{Gao2011, Abe2013, Jin2010}.
In CaH$_6$, a record-high $T_{\text{c}}$ of 220 to 235 K was reported at 150 GPa~\cite{Wang2012}.
In spite of these exciting theoretical results, it is fair to say that the system which has attracted most
attention resides on the recent discovered hydrogen
sulfides~\cite{Li2014,Duan2014,Duan2015,Errea2015,Li2016,Bernstein2015,Eigana2016}, especially after its
experimental observation of a 203 K $T_{\text{c}}$ based on direct transport measurement~\cite{Drozdov2015}.
Now, a consensus has been reached that the stable compound of hydrogen and sulfur at ambient pressure,
\textit{i.e.} H$_2$S, becomes unstable at high pressures.
The superconducting behavior at 200 GPa is very likely due to its decomposition to
H$_3$S~\cite{Duan2014,Duan2015,Li2016,Bernstein2015,Eigana2016}.
Comparisons of the \textit{ab initio} static enthalpy based on density-functional theory (DFT) calculations
(using the PBE functional)
have shown that the stoichiometry decomposition of H$_2$S to H$_3$S and S happens at 43 GPa, and H$_3$S remains
stable at least up to 300 GPa~\cite{Duan2015}.
Within this stability range of H$_3$S, an orthorhombic $Cccm$ structure dominates the group state between
43 and 112 GPa.
Then, the rhombohedral $R3m$ phase takes over and it remains most stable till 175 GPa.
After 175 GPa, the cubic $Im\bar{3}m$ structure becomes the most stable phase.
Calculations of the $T_{\text{c}}$ based on BCS theory using this $Im\bar{3}m$ structure of H$_3$S (D$_3$S),
in the mean time, show that this value can be as high as $\sim$200 K ($\sim$160 K).
In experiments, the $T_{\text{c}}$ of H$_3$S (D$_3$S) was measured to be $\sim$190 K ($\sim$150 K)
at 170 GPa~\cite{Drozdov2015}.
This agreement provides an excellent rationalization of the high $T_{\text{c}}$ phase with the
$Im\bar{3}m$ structure.
We note, however, that an isotope-dependence for the transition pressure to high $T_{\text{c}}$ exists in experiment,
\textit{i.e.} H$_3$S enters into the high $T_{\text{c}}$ region by $\sim$20 GPa earlier than
D$_3$S~\cite{Drozdov2015}.
Considering the fact that these theoretical studies are mostly based on static geometry optimizations,
the isotope-dependence of the experimental observation, especially the intrinsic anharmonic effects and the
nuclear quantum effects (NQEs), are rarely discussed.
Recently, using the stochastic self-consistent harmonic approximation (SSCHA) method for the treatment of nuclear
motion~\cite{Errea2013,Errea2014}, Errea \textit{et al.} showed that the phonon spectra of H$_3$S are highly
anharmonic and the anharmonic correction to the phonon spectra has a non-negligible influence on the values
of $T_{\text{c}}$~\cite{Errea2015}.
When taking into account of their contribution to the free-energy as the hydrogen (H) moves along
hydrogen-bond, they further demonstrated that the hydrogen-bond symmetrization in superconducting
H$_3$S has a strong quantum nature~\cite{Errea2016}.
Upon replacing H by its isotope deuterium (D), which has only one extra neutron and therefore possesses
less NQEs, the pressure with which the system transforms to the symmetric $Im\bar{3}m$ structure increases
by 12 GPa at 0 K.
Considering the fact that $T_{\text{c}}$ of the ansymmetric rhombohedral $R3m$ phase has a much lower
$T_{\text{c}}$ than the cubic $Im\bar{3}m$ phase, this quantum symmetrization of the hydrogen position along
the S-H$\cdots$S axis presents a clear picture for the isotope-dependence of the transition pressure with which
$T_{\text{c}}$ suddenly increases in H$_3$S and D$_3$S.
Their absolute values are still lower than the experimental observation by $\sim$50 GPa.
We note that this idea of quantum symmetrization was originally proposed in studies of high-pressure
H$_2$O~\cite{Goncharov1996,Loubeyre1999}, and \textit{ab initio} path-integral molecular dynamics (PIMD) simulations
have played an important role in their rationalization~\cite{Benoit1998}.
Confirmation of this picture and providing more physical insight to this phase transition by going to finite
$T$s, using methods like \textit{ab initio} PIMD, are necessary.
In addition to this, the PBE functional was used in descriptions of the electronic structure in Ref.~\onlinecite{Errea2016}\cite{PBE}.
It is well-known that these generalized gradient approximations (GGAs) to the exchange-correlation (XC) potential
often underestimate the proton transfer energy barrier in hydrogen-bonded systems~\cite{Cohen2011}.
A systematic study of the other exchange-correlation functional's influence on this symmetrization pressure
is highly desired.
Based on this consideration, we report in this manuscript a theoretical study on the H$_3$S
symmetrization, using \textit{ab initio} PIMD for the description of the NQEs at finite $T$s.
The accuracy of standard GGA functional was analyzed by comparing the potential energy surface (PES)
of the hydrogen along the S-H$\cdots$S axis with a hybrid functional.
At 90 and 160 K, our simulations show that NQEs influence the structures of H$_3$S and D$_3$S differently
and the interval when H$_3$S possesses the symmetric high $T_{\text{c}}$ structure while D$_3$S does not is in
agreement with, though their absolute values are lower than experiments.
This is consistent with a earlier theoretical study in Ref.~\onlinecite{Errea2016} where the SSCHA
method was used at 0 K.
The remaining discrepancy with experiments can be substantially decreased when the electronic structures are
described beyond PBE.
%
%
Our study presents a simple picture to interpret the isotope-dependent of $T_{\text{c}}$, rationalizes the
remaining discrepancy with experimental using higher-level structures, and emphasizes the quantum
nature of the high-pressure hydrogen sulfide system.
The paper is organized as follows.
The computation details are given in Sec. II.
In Sec. III, we presents our results on the quantum nature of the S-H$\cdots$S axis symmetrization, as
well as an analysis of the influence of functionals.
We draw our conclusions in Sec IV.

\section{II. Computational Details}
Our simulations were preformed using the Vienna \textit{ab initio} Simulation Package (VASP)
code~\cite{vasp1,vasp2}, along with our own implementation of the PIMD
method~\cite{Chen2013,Chen2014,Feng2015,Guo2016,Chen2015,Feng2016}.
DFT was used to describe the electronic structure ``on-the-fly'' as the path of the nuclei propagates.
Projector augmented wave (PAW) potentials along with a 700 eV energy cutoff were employed for the expansion
of the electronic wave functions~\cite{paw1,paw2}.
The Perdew-Burke-Ernzerhof (PBE) functional was used to describe the electronic exchange-correlation interaction~\cite{PBE}.
With a supercell containing 96 atoms and a Monkhorst-Pack k-point mesh of spacing 2$\pi\times0.04$\AA\text{}$^{-1}$
to sample the Brillouin zone, we performed \textit{ab initio} PIMD simulations at 90 and 160 K.
The Andersen thermostat was chosen to control the temperature of the $NVT$ ensemble~\cite{Andersen1980}, in which
the atomic velocities were periodically randomized with respect to the Maxwellian distribution every 60 fs.
%

%
%
For the reported results, 16 beads were used.
%
All statistics of the bond length were obtained using the centroid of the path, and we note that using the
bond lengths in the individual beads gives the same results.
The PBE exchange-correlation functional used in the PIMD simulations suffers from self-interaction errors, which
may induce a substantial underestimation of the transition-state energy~\cite{Cohen2011}.
We investigate this effect by focusing on the transition barrier and transition distance of the hydrogen atoms
as they move along the S-H$\cdots$S axis between their equivalent positions, and comparing the values
obtained using the PBE and the hybrid HSE06 functional~\cite{hse03,hse06}.
In determining these energy barriers, the climbing image nudged elastic band (cNEB) method was used~\cite{cNEB},
in which all force components that perpendicular to the tangent of the reaction path were reduced to less than
0.01 eV/\AA\text{}.

\section{III. Results and Discussions}

\subsection{IIIa. Quantum Nature of the Phase-Transition}
\begin{figure}[!ht]
\centering
\includegraphics[width=0.43\textwidth]{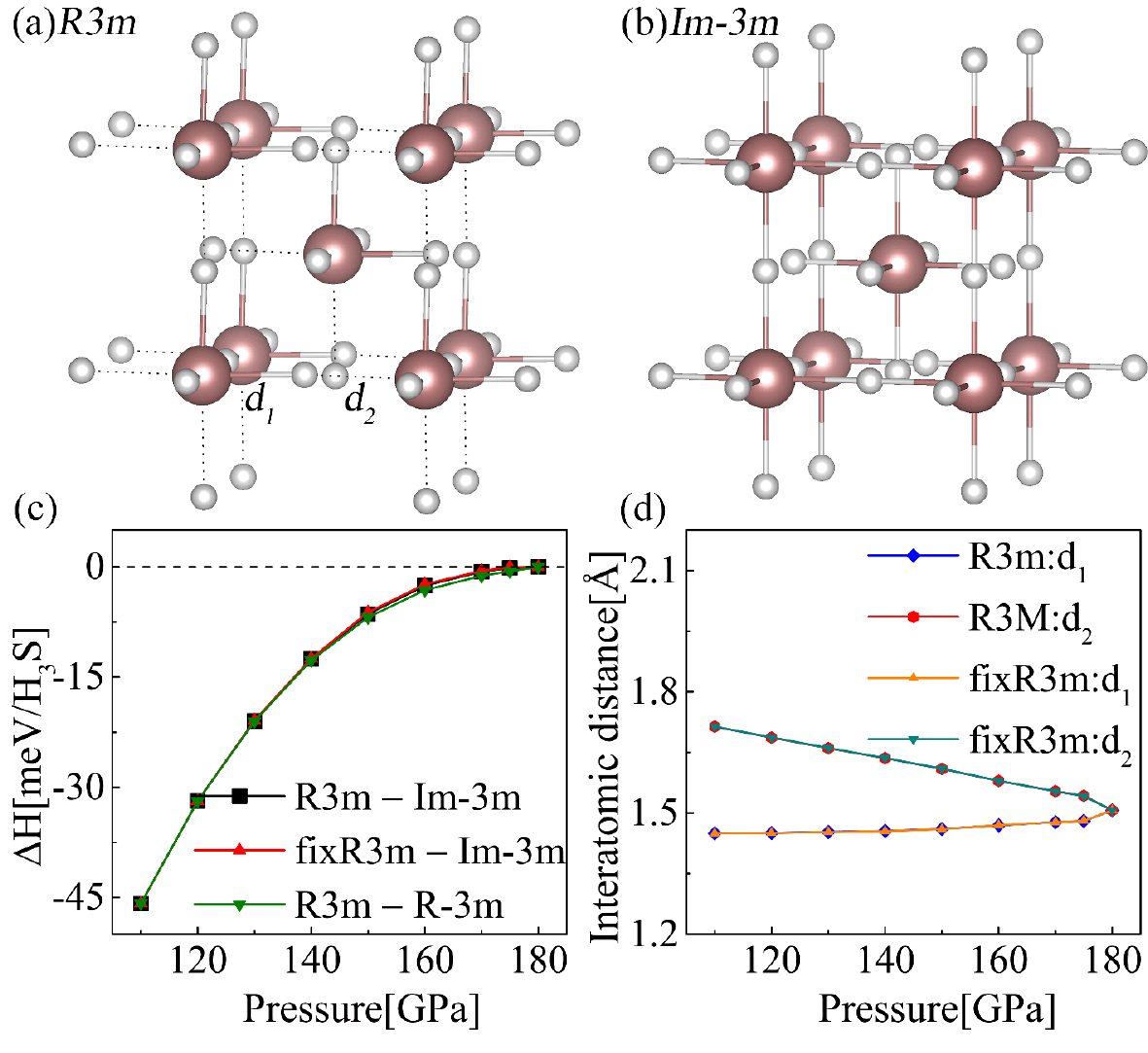}
\caption{\label{FIG1}
Static crystal structures of the $R3m$ (panel a) and $Im\bar{3}m$ (panel b) phases.
A slightly distorted bcc lattice is used for the $R3m$ phase and the conventional bcc lattice is used for $Im\bar{3}m$.
$d_1$ and $d_2$ are the distance between the H atom and its two nearest neighboring S atoms.
In $R3m$, these two values are not equal. In $Im\bar{3}m$, they are.
%
In c), we compare the enthalpy differences of the $R3m$ and $Im\bar{3}m$ phases by imposing a cubic lattice for $R3m$ (fix$R3m$) or a rhombohedral lattice for the symmetric structure ($R\bar{3}m$).
In d), we monitor the evolution of $d_1$ and $d_2$ using the cubic and the rhombohedral cells.
These two panels show that imposing the rhombohedral symmetry has a negligible influence on the symmetrization of
the hydrogen bonds.
%
Summing up these four panels, we would say that the main structural change upon transforming from $R3m$
to $Im\bar{3}m$ lies on symmetrization of the hydrogen bonds.
}
\end{figure}

We start by looking at the structures of the rhombohedral $R3m$ phase and the cubic $Im\bar{3}m$ phase at the
static level.
Although being rhombohedral, $R3m$ is very close to cubic symmetry.
This can be evidenced by looking at the rhombohedral angle of $R3m$, which is 109.5142, 109.5476,
and 109.5592, respectively
at 140 GPa, 150 GPa, and 160 GPa, in comparison with the value of 109.47$^\circ$ for a perfect bcc lattice.
Therefore, we visualize the difference between the structures of these two phases in Fig.~\ref{FIG1} a)
and b), using a slightly distorted bcc cell of the $R3m$ phase and the conventional bcc cell of the
cubic $Im\bar{3}m$ phase.
The S atoms stay on a slightly distorted bcc lattice in the $R3m$ phase and the conventional bcc lattice
in the $Im\bar{3}m$ phase.
The difference between these two lattices is visually indistinguishable.
The H atoms, however, show very different behaviors.
In the $R3m$ phase, the H atoms stay asymmetrically along the S-H$\cdots$S axis, being closer to one S atom.
If we label the covalent bond length S-H as $d_1$ and the distance between the H atom and its next nearest S atom
as $d_2$, these two distances are unequal.
In the $Im\bar{3}m$ phase, however, they are symmetric, with $d_1=d_2$.
Therefore, the main change happened during the phase transition from the $R3m$ phase to the $Im\bar{3}m$ phase,
lies on the symmetrization of S-H$\cdots$S, instead of the rhombohedral to cubic lattice evolution.
%
This can be clearly seen in Fig.~\ref{FIG1} c) and d), where we show that using a rhombohedral lattice for
the symmetric structure ($R\bar{3}m$) or a cubic lattice for $R3m$ (fix$R3m$) has negligible effect on enthalpy differences of these two phases, and by imposing rhombohedral or cubic lattice for $R3m$,
the symmetrization happens at very close pressures.
In Ref.~\onlinecite{Errea2016}, a cubic cell has been used in discussion of the phase transition in.
Here, we adopt a rhombohedral cell in discussions of the classical and quantum nuclear effects at
finite $T$s.
By comparing with their results, one will see that the quantum feature of hydrogen-bond symmetrization is
robust with respect to this choice of the simulation cell.
%

\begin{figure}[!ht]
\centering
\includegraphics[width=0.5\textwidth]{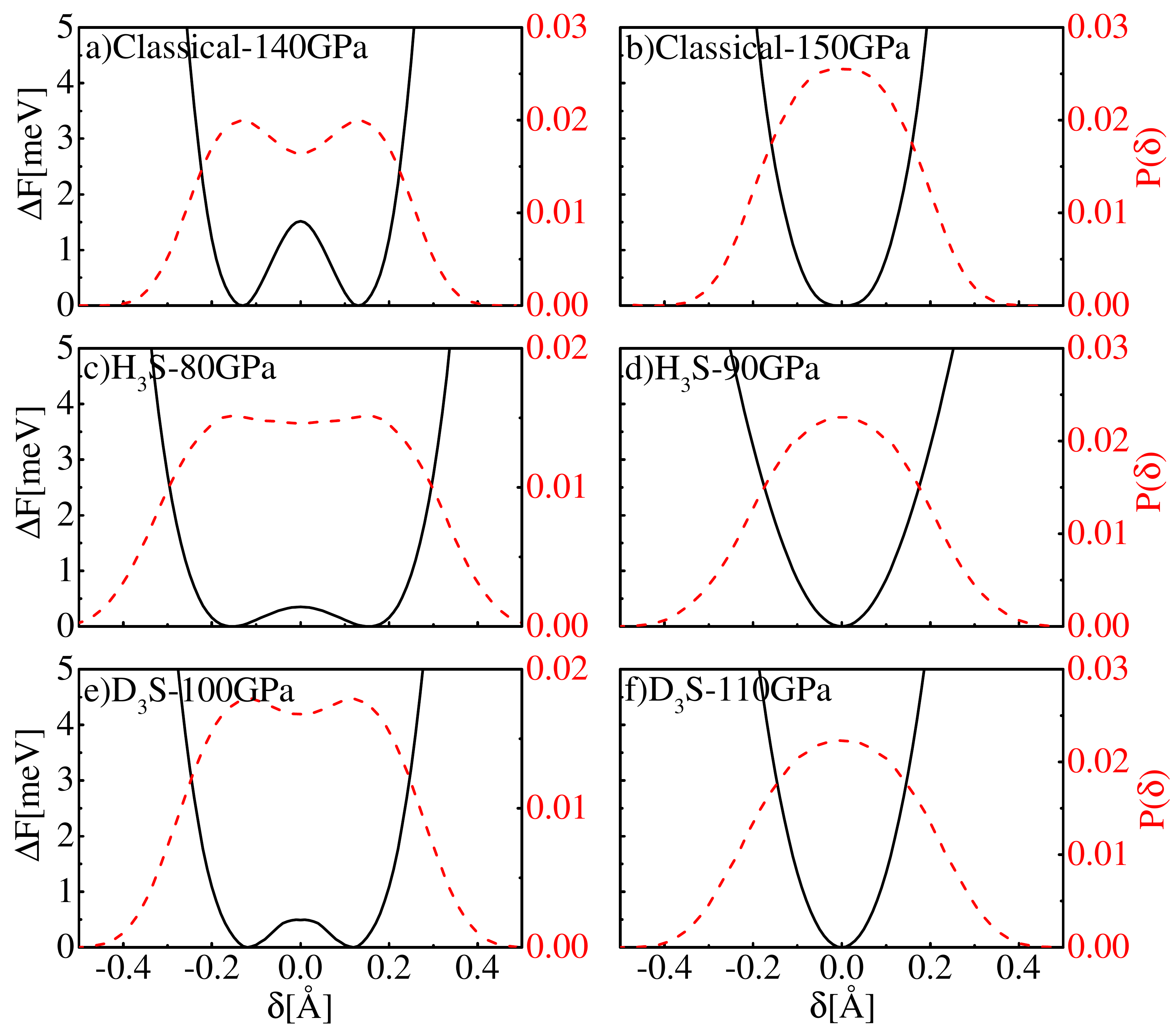}
\caption{\label{FIG2}
Symmetrization of the hydrogen bonds.
Using the H transfer coordinate (defined as $\delta = d_1-d_2$), the probability distribution $P(\delta)$
and the free-energy profile $\Delta F(\delta)$ are plotted to the left and right scales based on
\textit{ab initio} MD and \textit{ab initio} PIMD simulations at 90 K.
Panels a) and b) correspond to results using classical nuclei, where the upper bound for the asymmetric
structure and lower bond for the symmetric structure are reported within numerical resolution of 10 GPa.
Panels c) and d) correspond to results obtained using quantum nuclei for D$_3$S.
And panels e) and f) correspond to results obtained using quantum nuclei for H$_3$S.
}
\end{figure}

Now we include the nuclear statistical effects at the classical level.
This is done by carrying out the \textit{ab initio} molecular dynamics (MD) simulations at 90 K for a supercell
of the $R3m$ phase.
We define a H transfer coordinate $\delta=d_1-d_2$.
When it is zero, the hydrogen bond is symmetric and the H atom is equally shared by the two S atoms.
When it is large in magnitude on both positive and negative sides, the H belongs to one S atom.
The results are shown in Fig.~\ref{FIG2} a) and b).
It is clear that although the hydrogen bond symmetrization happens at $\sim$170 GPa at the static level,
this transition pressure is substantially decreased at a finite $T$ upon including the nuclear classical
statistical effects.
At 140 GPa, the probability distribution ($P(\delta)$) has two clear peaks at $\delta$ equals
$\pm$0.2 \AA\text{}, indicating that the hydrogen bonds are still asymmetric.
At 150 GPa, it has a single peak at $\delta=0$, a clear indication of hydrogen bond symmetrization.
Therefore, the symmetrization of the hydrogen bond happens in between 140 and 150 GPa, when the nuclear
classical statistical effects are included at 90 K.
This symmetrization is also confirmed by the corresponding free-energy profile calculated using
$\Delta F(\delta)=-k_{\text{B}} T \text{ln} P(\delta)$, where the single- and double-valley feature of
the profiles tells us where the H atoms want to stay.
In comparison with the static results, with classical nuclei, the symmetrization moves toward a lower
pressure at a finite $T$.

\begin{figure}[!ht]
\centering
\includegraphics[width=0.45\textwidth]{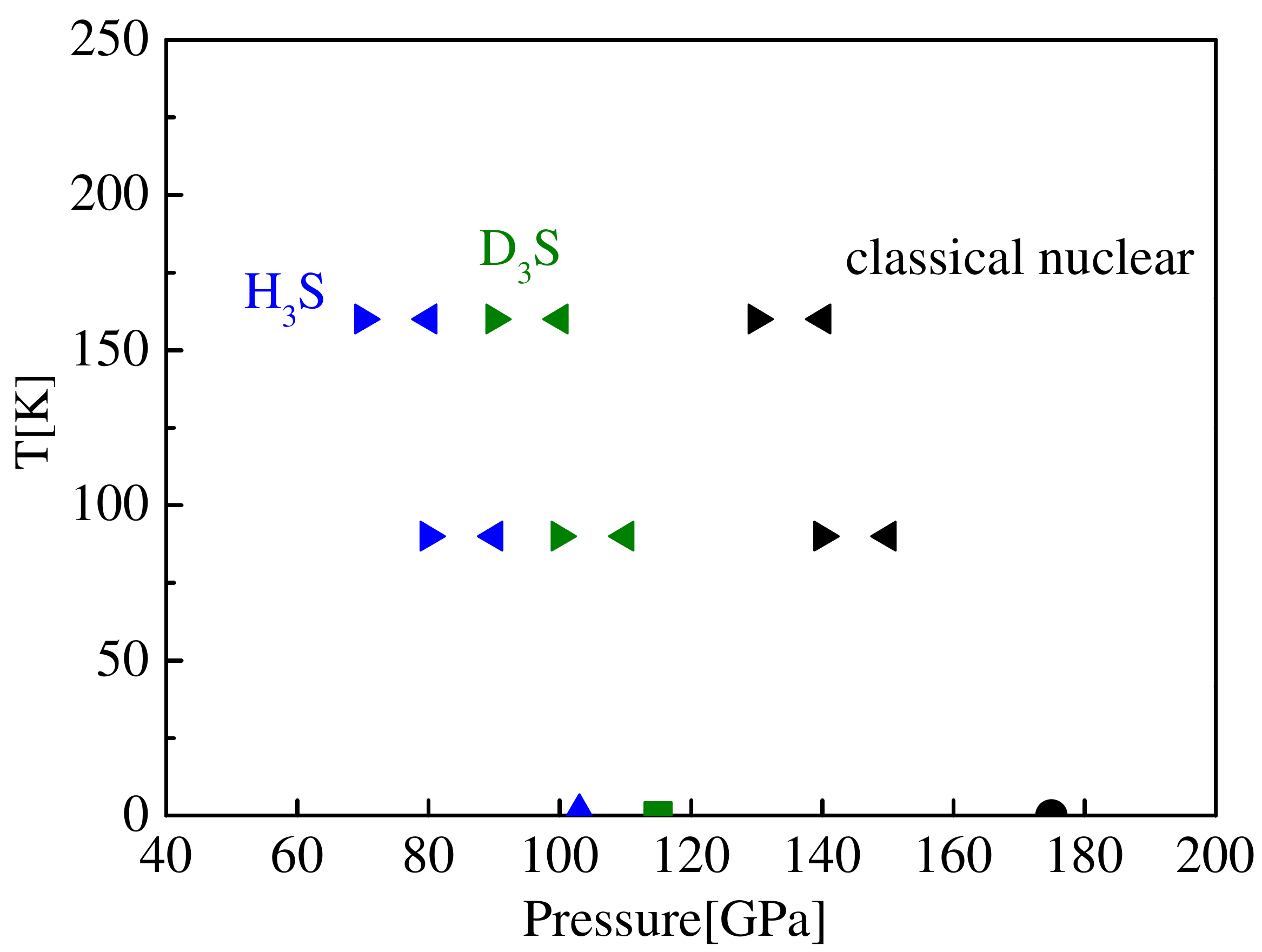}
\caption{\label{FIG3}
$T$-dependence of the hydrogen bond symmetrization using classical nuclei, and quantum nuclei for D$_3$S
and H$_3$S.
The upper (lower) bounds of the asymmetric (symmetric) hydrogen bonds are labelled using left-(right-)oriented
triangles respectively.
Two series of \textit{ab initio} MD and \textit{ab initio} PIMD simulations are reported.
The results from Ref.~\onlinecite{Errea2016} for classical nuclei and quantum nuclei (D$_3$S and H$_3$S)
at 0 K were labelled using blue upward triangle, green half square, and black half circle.
}
\end{figure}

Then we include the NQEs for D at the same $T$.
The structure is analyzed in a similar way, using the centroid of the path-integral in Fig.~\ref{FIG2} c) and d).
At 90 GPa, $P(\delta)$ ($\Delta F(\delta)$) shows a double-peak (double-well) structure, indicating that the hydrogen
bonds are still not symmetric.
At 100 GPa, however, the single-peak (single-well) structure already appeared, meaning that the hydrogen bond
symmetrization happened.
Therefore, the transition from asymmetrized to symmetrized hydrogen bond happens between 90 and 100 GPa in D$_3$S.
In H$_3$S, on the other hand, this symmtrization happens between 80 and 90 GPa (Fig.~\ref{FIG2} e) and f)).
We note that these values for H$_3$S and D$_3$S are lower than the transition pressures of 103 and 115 GPa when
quantum nuclei are used at 0 K.
To investigate in more detail the $T$-dependence of this transition pressure, we carried out a separate series
of \textit{ab initio} MD and \textit{ab initio} PIMD simulations at 160 K.
Using right- and left- solid triangles, we label the upper and lower boundary of asymmetric and symmetric structures
in Fig.~\ref{FIG3}, along with results discussed earlier for the 90 K simulations.
The results reported for classical nuclei and quantum nuclei (both D$_3$S and H$_3$S) at 0 K
in Ref.~\onlinecite{Errea2016} were also labelled using diamond, solid circle, and open circle respectively.
From these figure, it is clear that while being consistent with the isotope-dependence of the symmetrization
pressure as reported in Ref.~\onlinecite{Errea2016}, our simulations also show that the symmetrization happens
at lower pressures upon increasing $T$.
This is also in agreement with a series of earlier hydrogen bond symmetrization study of ice under
pressure~\cite{Goncharov1996,Loubeyre1999,Benoit1998,XZLi2010}.
Considering the fact that symmetric $Im\bar{3}m$ phase exhibits a much higher $T_{\text{c}}$ than the
asymmetric $R3m$ phase when calculations based on BCS theory is used~\cite{Duan2014}, this presents
a clear picture for the isotope-dependence of the transition pressure of H$_3$S and D$_3$S to the high
$T_{\text{c}}$ phase~\cite{Drozdov2015}.
In comparison with the results obtained using classical nuclei, the hydrogen bonds in H$_3$S are
strongly quantum in nature.

\subsection{IIIb. Deficiency of PBE Functional}

In spite of the qualitatively excellent agreement between the theoretical and experimental results, as
reported above and in Ref.~\onlinecite{Errea2016}, a large discrepancy still exists.
The predicted symmetrization happens at a pressure $\sim$60 GPa lower than the experimental observation.
Considering the fact that the proton-transfer energy barriers and the distance of proton transfer play
an important role for such a symmetrization~\cite{XZLi2010}, and the PBE functional used here and
in Ref.~\onlinecite{Errea2016} is well-known to underestimate the transition state (TS) energy in
description of the chemical reactions~\cite{Cohen2011},
it is reasonable to expect that the deficiency of the PBE functional plays a key role on the underestimation
of this transition pressure.
Based on this consideration, we investigate the functional-dependence of the symmetrization pressure.

\begin{figure}[!ht]
\centering
\includegraphics[width=0.43\textwidth]{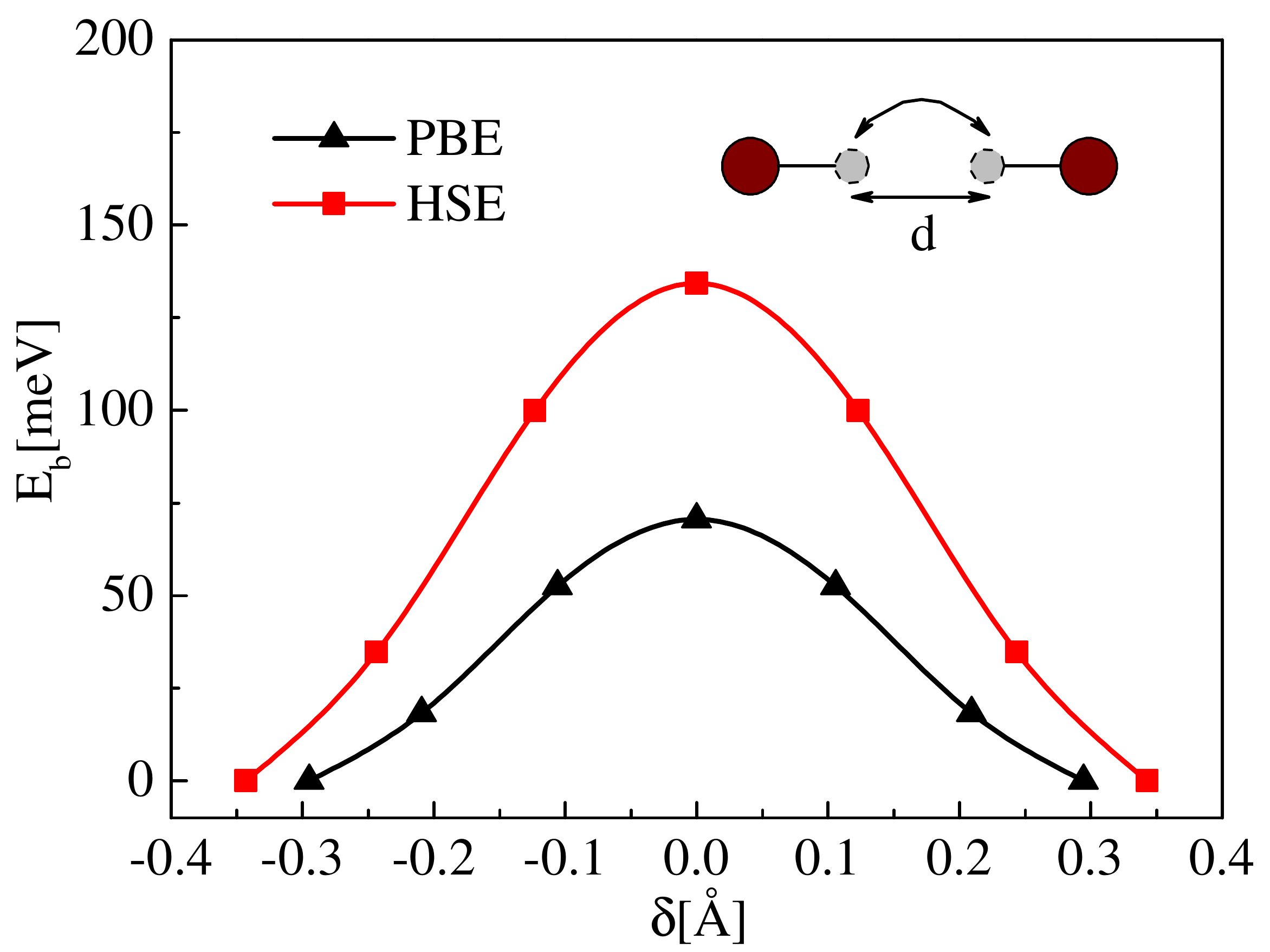}
\caption{\label{FIG4}
TS searching based on the cNEB method for the proton transfer process.
The red squares are obtained using HSE06.
The black triangles are obtained using PBE.
Upon including partial exact exchange, the proton transfer barrier and the proton transfer distance increase.
This is a clear indication that self-interaction substantially underestimates these quantities, and
consequently underestimates the symmetrization pressure.
}
\end{figure}
For periodic system, the easiest way to check this underestimation of the TS energy, mostly due to the
delocalization error~\cite{Cohen2011}, is to compare the results obtained from the hybrid-functional
calculations with those obtained from the standard local-density approximation (LAD) or generalized-gradient
approximation (GGA) calculations.
However, considering the fact that the computational cost of performing a hybrid-functional based PIMD
simulation is beyond what we can afford, we carried out such an analysis at the static level.
The cNEB method is used to find the TS for the H atom transfer from one S atom to the other.
The energy-profiles are shown in Fig.~\ref{FIG4} for the simulation at 100 GPa.
From this figure, it is clear that the PBE functional has underestimated the proton transfer energy barrier
by 50\% in comparison with the HSE06 based results.
Considering the fact that partial exact-exchanges are included in this hybrid-functional, which helps to
cure the artificial electron charge delocalization induced by self-interaction~\cite{Cohen2011}, this  is
a clear indication that PBE has underestimated the proton transfer energy barrier.
Consequently, when the \textit{ab initio} PIMD simulations are performed using electronic structures provided
by the DFT calculations using PBE functional, a systematic underestimation of the symmetrization pressure occurs.
We note that a similar underestimation of the transition pressure also happens in the molecular liquid
dissociation to the atomic liquid phase in high-pressure hydrogen, as pointed out in Ref.~\onlinecite{Chen2014}.
Upon including more exact-exchange interaction in this HSE-based functional, this underestimation is even
more serious.
Therefore, we believe that the right answer for proton transfer energy barrier must lie somewhere much higher than
the PBE result.
A quantitative determination of this value requires more sophisticated electronic structure methods, which is
beyond the scope of the present paper.
%
%

\begin{figure}[!ht]
\centering
\includegraphics[width=0.43\textwidth]{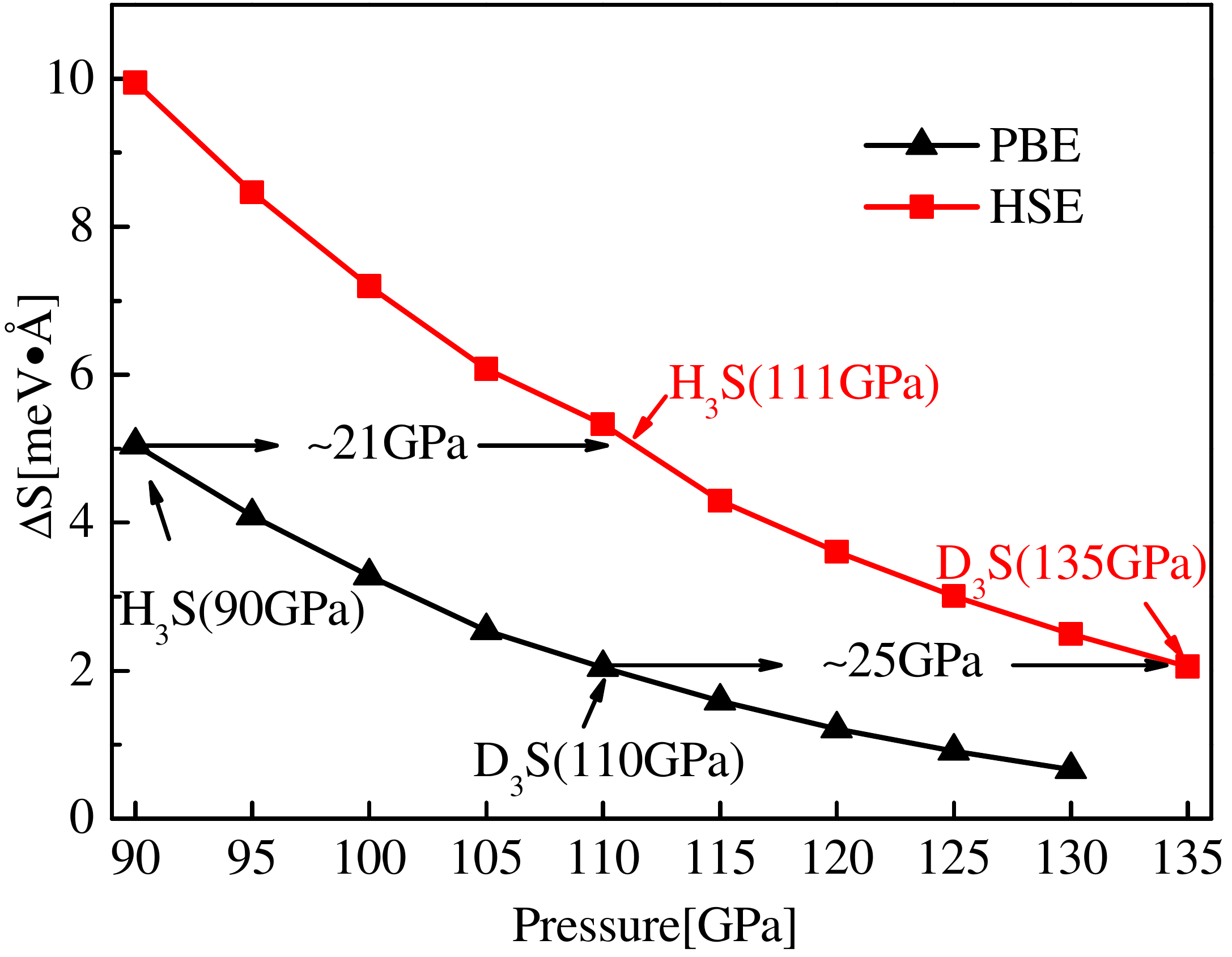}
\caption{\label{FIG5}
Estimation of the symmetrization pressure, using $\Delta S = d\times\Delta E$, as the descriptor.
The upper horizontal line at $\Delta S\sim5$ indicates the situation when H$_3$S symmetrization is confirmed.
The upper horizontal line at $\Delta S\sim2$ indicates the situation when D$_3$S symmetrization is confirmed.
Based on this extrapolation, we estimate that using the HSE function will increase the symmetrization pressure
of H$_3$S (D$_3$S) by 21 (25) GPa, toward the experimental results.
}
\end{figure}
Finally, despite hybrid-functional based \textit{ab initio} PIMD simulations are beyond the
computational cost we can afford, due to the fact that extend of proton delocalization is sensitive to the
area below the proton transfer energy profile, an estimation of the symmetrization pressure can be
provided.
An estimator $\Delta S$ is defined as ($d\times\Delta E$)/2 and plotted as a function of pressure in Fig.~\ref{FIG5}.
The definition of $d$ is given in the inset of Fig.~\ref{FIG4}.
At each pressure, this $\Delta S$ is obtained by two steps.
First, an enthalpy-based geometry optimization is carried out using DFT calculations with PBE and HSE06
functionals.
The difference between the two symmetrized structures with the H atom belonging to each S gives us $d$.
Then, $\Delta E$ is calculated using the cNEB method, again based on DFT calculations with PBE and HSE06
functionals.
The results are shown in Fig.~\ref{FIG5}.
From Fig.~\ref{FIG4}, we know that the difference between the values of $d$ is small and the difference between
values of $\Delta E$ is large between DFT calculations with the PBE and the HSE06 functionals.
Consequently, at each pressure, the value $\Delta S$ is much larger in the HSE06-based DFT calculations compared
with the PBE-based DFT calculations.
The horizontal dashed line indicates the value of $\Delta S$ at which H$_3$S symmetrizes and the horizontal solid
line indicates the value of $\Delta S$ at which D$_3$S symmetrizes.
Therefore, an estimation of the symmetrization pressure in the HSE06 functional based PIMD simulations would be
110 GPa for H$_3$S and 135 GPa for D$_3$S at 90 K.
Compared with the PBE functional based PIMD simulations, these values clearly move toward experimental observation.
Again, we note that a better agreement between theory and experiment must require more accurate theoretical methods
on the electronic structure level, with NQEs also accurately addressed.
This is beyond the scope of most theoretical groups.
We focus more on pointing out the deficiency of the PBE functional, so that further theoretical studies could
be aware of in such simulations.

\section{Conclusions}
Based on \textit{ab initio} MD and \textit{ab initio} PIMD simulations, we systematically investigate the
influence of nuclear statistical effects on the symmetrization of H$_3$S and D$_3$S with classical and
quantum nuclei at finite $T$s.
The accuracy of standard GGA functional was analyzed by comparing the PES of the hydrogen along the
S-H$\cdots$S axis with the hybrid ones.
Our simulations show that NQEs influence the structures of H$_3$S and D$_3$S differently
and the interval when H$_3$S possesses the symmetric high $T_{\text{c}}$ structure while D$_3$S does not is in
agreement with, though their absolute values are lower than the experimental observations.
These results at 90 and 160 K are consistent with a earlier theoretical study in Ref.~\onlinecite{Errea2016}
where the SSCHA method was used at 0 K.
The remaining discrepancy with experiments can be substantially decreased when the hybrid-functional is used.
%
%
This study presents a simply picture to interpret the isotope-dependent of $T_{\text{c}}$.
In the meantime, it also rationalizes the remaining discrepancy with experiments by pointing out the deficiency of
the PBE functional, and emphasizes the quantum nature of the high-pressure hydrogen sulfide system.

\begin{acknowledgements}
Y.Y., Y.X.F., and X.Z.L. are supported by the National Basic Research Programs of China under Grand
No. 2013CB934600, and the National Science Foundation of China under Grant Nos 11275008, 11422431.
Y.Y. and L.F.B. are supported by the Open-Lab program (Project No. 12ZS01) of the Key Laboratory of
Nanodevices and Applications, Suzhou Institute of Nano-Tech and Nano-Bionics (SINANO), Chinese Academy of
Sciences.
The computational resources were provided by the supercomputer TianHe-1A in Tianjin, China.
\end{acknowledgements}

 \bibliography{ref}

\end{document}